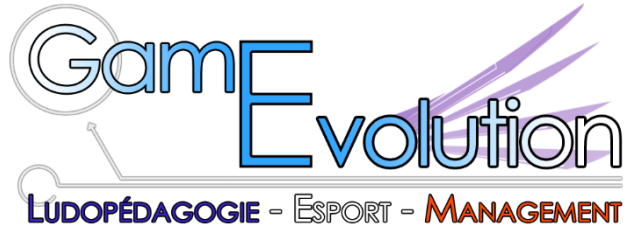

8e Colloque International *Game Evolution*

# La gamification de l'informatique au service de la formation en management des systèmes d'information

**Yann Goetgheluck**
IAE Paris-Est
yann.goetgheluck@etu.u-pec.fr

**Sarah Mernit**
IAE Paris-Est
sarah.mernit@etu.u-pec.fr

**Julie Pereira**
IAE Paris-Est
julie3.pereira@etu.u-pec.fr

**Résumé :**

Cet article examine l'intégration des compétitions informatiques, notamment le *Capture The Flag*, dans une formation en management des systèmes d'information pour combler les lacunes de compétences, en particulier dans le domaine de la cybersécurité. Une équipe CTF pédagogique a été mise en place à l'IAE Paris-Est avec pour objectif de développer les compétences des étudiants. Des ateliers, des défis et des événements ont été organisés pour les familiariser avec les CTF et leur offrir un accompagnement adapté à leur niveau. Les résultats préliminaires montrent l'importance des *soft skills* dans l'amélioration des compétences en cybersécurité. L'équipe CTF pédagogique continue d'expérimenter et d'évaluer ces méthodes pour améliorer l'accessibilité et l'efficacité de la formation en cybersécurité.

**Mots-clefs :**

Cybersécurité ; Capture de drapeaux ; *Hacking* éthique ; Management des systèmes d'information ; Ludopédagogie ; Gamification ; Compétences comportementales

# Gamification of IT for training in information systems management

**Abstract:**

This article examines the integration of IT competitions, in particular Capture The Flag, into an information systems management course to fill skills gaps, particularly in the field of cybersecurity. An educational CTF team has been set up at IAE Paris-Est with the aim of developing students' skills. Workshops, challenges, and events have been organised to familiarise them with the CTFs and offer them support adapted to their level. Preliminary results show the importance of soft skills in improving cybersecurity skills. The CTF pedagogical team is continuing to experiment with and evaluate these methods to improve the accessibility and effectiveness of cybersecurity training.

## 1. Introduction

L'informatique, un domaine vaste et complexe, trouve une dimension ludique à travers les compétitions informatiques. En particulier les compétitions *Capture The Flag* (CTF), des événements d'informatique dont le but des participants est de trouver une suite de caractères spécifiques nommé « *flag* ». Ils permettent ainsi de rendre ce domaine plus accessible et de combler les lacunes de compétences, surtout dans le domaine de la cybersécurité (Cobb, 2016, p.1). Notre initiative vise à exploiter cette dynamique en accentuant l'accessibilité aux compétences informatiques par le biais d'une approche ludopédagogique.[1] Dans cette optique, nous examinons comment un projet de formation, centré sur les compétitions CTF, contribue à résoudre les défis éducatifs liés à l'acquisition des compétences complexes du *hacking*. Cela en utilisant au maximum l'aspect de gamification des CTF. Ce qui « consiste à recourir aux mécaniques du jeu dans le cadre de plusieurs activités (professionnelles, mais aussi de la vie quotidienne) à l'origine non-ludiques afin de changer le comportement d'une personne ou d'un groupe de personnes » (Lépinard & Vandangeon-Derumez, 2019). Cela est très utile pour stimuler l'engagement et la motivation des apprenants dans des activités d'interaction humain-machine (Marache-Francisco & Brangier, 2015). Notamment dans le cadre de la formation en management des systèmes d'information. Nous avons lancé ce projet de recherche dans le but de répondre à deux questions fondamentales. En premier lieu, comment pouvons-nous aider les étudiants en Management des Systèmes d'Information à améliorer leurs compétences techniques de manière efficace ? Pour cela, nous avons opté pour l'approche des CTF. Nous examinerons donc les premiers résultats après avoir établi l'importance de ce travail et mis en place les mesures nécessaires pour expérimenter les CTF pédagogiques. De plus, nous cherchons à répondre à la question suivante : quels sont les éléments essentiels pour progresser efficacement dans les compétitions informatiques ? Pour ce faire, nous identifierons les éléments que nous avons observés et expliquerons leur impact sur la progression.

## 2. Pourquoi choisir les CTF ?

Les CTF, cyber-compétitions « *Captures The Flag* » sont des événements permettant d'enseigner et de tester les compétences en matière d'informatique, de sûreté de l'information et de sécurité des réseaux grâce à la ludification et au développement ciblé des compétences.

---

[1] La ludopédagogie est « l'ensemble des activités ludiques déployées par un formateur pour développer des apprentissages chez autrui dans un contexte pédagogique formel ou non formel » (Lépinard & Vandangeon-Derumez, 2019)

L'objectif de chaque CTF est d'obtenir une information précise appelée *flag*. Balon et Baggili (2023) expliquent que ces cyber-compétitions sont nées à la suite des premiers concours de programmation et les conférences sur le piratage dans les années 1970 et 1980. Les compétitions de défense et des tests de pénétration dans les années 1990 et 2000. Aujourd'hui, les cyber-compétitions visent à attirer et à éduquer des publics plus jeunes et plus diversifiés. Sa gamification permet d'améliorer les résultats d'apprentissage, d'engagement, et même la motivation des enseignants en sécurité des systèmes d'information (Marache-Francisco & Brangier, 2015).

Selon Cole (2022, p. 470), il existe de nombreux types de CTF. Tout simplement des *Quizz,* le type de niveau le plus basique, permettant un format simple de questions et de réponses. Ensuite, le type *Jeopardy*, le plus courant, c'est un ensemble de défis fournis par les organisateurs du concours aux concurrents contre une machine. Le type *Attack-Defense,* est une bataille ou chaque équipe dispose de son réseau avec des vulnérabilités s'attaquant mutuellement. Pour finir, *King of the Hill*, qui ressemble au type *Attack-Defense*, la seule différence réside dans l'unique ressource, réseau à attaquer, dont il faut prendre le contrôle et le garder le plus longtemps possible. Ce large éventail de type de défis associés aux divers domaines de connaissances permet de toucher à tous les aspects de l'informatique dans des situations réalistes, complexes et complètes. Ces cyber-compétitions ont comme leur nom l'indique un intérêt non-négligeable pour amplifier l'intérêt des participants, en effet, la compétition est un outil très utile. Cela permet de stimuler l'émergence de rivalité saine et de se pousser les uns et les autres vers le haut. Nonobstant le fait que ces compétitions se font la majorité du temps en équipe, cette atmosphère de compétition, de tension apporte un réel environnement extrême pour le travail en équipe. Nous sommes dans une situation proche de la crise, situation à laquelle des responsables de système d'information sont voués à connaître. Et cela avec des tiers qui tentent soit d'aller plus vite, soit de faire mieux que vous. On peut parfois comparer la situation des CTF à celle d'une cyber-attaque subit par un responsable de la sécurité des systèmes d'information (RSSI) et son équipe. Et c'est exactement ce que nous recherchons.

En plus de cette mise en situation, nous avons lors de ce projet, réalisé l'impact des récompenses. Au lieu de sanctionner celles et ceux qui n'arrivent pas des tâches techniques et/ou organisationnelles, puisque c'est un travail d'équipe, avec de notes démoralisantes, il est préférable de récompenser les personnes qui réussissent ces tâches. Cela maintient la morale des équipes et renforce la rivalité, ce qui permet d'avoir des résultats plus importants. « L'activation dans notre cerveau du circuit de la récompense aboutit à la libération de dopamine, le messager chimique du plaisir » (Maheu, 2023). Nous avons pu établir, qu'en moyenne, il y a une différence de 30% de volontaire entre un événement qui promet des récompenses et un qui n'en promet pas. Il faut néanmoins utiliser ce levier avec parcimonie, afin que les récompenses ne soient pas le symbole des différences de niveau. Il ne faut pas que cela devienne un moyen de clivage avec d'un côté les récompensés et d'un autre les non-récompensés.

Par ailleurs, un aspect des CTF qui est primordial et que nous avons évoqué précédemment, le travail d'équipe, représente aussi un élément en faveur de cet outil de formation. Savoir travailler en équipe est un besoin nécessaire dans toute organisation[2], et tout particulièrement dans le domaine de la sécurité informatique où la collaboration efficace peut faire la différence entre une cyber-attaque maîtrisée et une catastrophe totale. Les CTF offrent une plateforme idéale pour développer et mettre à l'épreuve les compétences en travail d'équipe, en mettant les participants au défi de collaborer étroitement pour résoudre des problèmes complexes dans un laps de temps limité. En travaillant ensemble, les membres d'une équipe peuvent combiner leurs forces, exploiter les compétences individuelles et partager les connaissances pour atteindre un objectif commun : la capture du drapeau. Cette expérience de travail collaboratif sous pression reflète fidèlement les défis rencontrés dans le monde réel, où la coordination et la communication efficaces sont essentielles pour répondre rapidement et efficacement aux menaces de sécurité. En outre, les CTF favorisent également le développement de compétences interpersonnelles telles que la confiance, le respect mutuel et la gestion des conflits, qui sont toutes des compétences essentielles pour réussir dans le domaine de la sécurité informatique. Ce qui nous amène, dans ce travail, à ne pas nous focaliser sur les compétences techniques, mais aussi sur ces compétences interpersonnelles.

La grande variété d'application des CTF, l'aspect compétition, le côté ludique, l'utilisation des récompenses et la double compétence, technique et humaine. Ils représentent les principaux éléments en faveur de l'utilisation des CTF dans l'optique d'une montée en compétences en sécurité informatique. En effet, les différents types de CTF permettent de se concentrer sur des aspects précis. Par exemple, *Quizz* ou *Jeopardy* permet de voir les connaissances brutes des équipes, *Attack-Defense* et *King of the Hill* permet de se concentrer sur la capacité d'adaptation des participant et de collaboration. La liberté que nous offre ce terrain de jeu que représentent les CTF, nous semble être un élément qui peut parfois faire défaut aux travaux de groupe dirigés. Et ajoutons à cela que nous avons rapidement pu remarquer l'intérêt de la compétition et des récompenses lors des premiers événements organisé par l'équipe CTF pédagogique.

---

[2] « *Une organisation est un système social délibérément mis en place pour atteindre des objectifs spécifiques en utilisant des ressources humaines, matérielles et financières, et caractérisé par une structure hiérarchique, des processus de prise de décision et des mécanismes de coordination.* » (Robbins & Coulter, 2020).

## 3. Intégration des CTF pour la formation en MSI

Contrairement à la proposition d'un unique cours ludifié de Arduin & Costé (2022, p.7 et suivantes) nous avons décidé de créer une équipe CTF pédagogique (Figure 1 et Tableau 1), au sein de l'école universitaire de management de l'IAE Paris-Est, et soutenue par ses enseignants. Ce projet entre dans le cadre de la montée en compétences des étudiantes et étudiants du Parcours Informatique et Management. Au cours de cette année universitaire 2023-2024, notre équipe pédagogique CTF a pour but de faire découvrir, accompagner et aider les membres de la formation à progresser sur l'aspect technique de la cybersécurité. L'idée de ce projet de recherche-action (Baskervill & Meyers, 2004 ; Charliès *et al.*, 2018) provient d'un enseignant souhaitant trouver la réponse à la question suivante : quelles compétences en cybersécurité devraient posséder les étudiantes et étudiants en Management des Systèmes d'Information ? Tout cela a bien entendu comme objectif premier d'aider les étudiantes et étudiants à obtenir une meilleure compréhension des systèmes d'information[3] de manière ludique. Nous avons choisi l'utilisation des CTF pour les raisons évoquées dans la partie précédente. Et pour réaliser ce travail nous avons une équipe de dix étudiantes et étudiants volontaires, qui a été formée dans le cadre d'un projet universitaire, composé de tous les niveaux présents dans une formation universitaire afin de pouvoir s'adapter, sur tous les sujets et tous les points de vue possible. Dans l'optique d'obtenir une équipe compétente, le premier objectif à court terme a été de s'auto former. Pour cela, nous avons commencé par la résolution des défis du module « *Beginner* » sur la plateforme 247CTF[4]. Par la suite, nous avons continué à développer les compétences des membres de l'équipe. En réalisant un grand nombre de défis de niveaux « Débutant », « Facile » et « Intermédiaire » sur la plateforme Ozint[5]. Enfin, nous avons terminé notre préparation par la réalisation du défi « Bandit » sur la plateforme OverTheWire.[6]. Ainsi, grâce à ce socle de connaissances, chaque membre de l'équipe est maintenant armé pour aider et soutenir la progression de chaque membre du parcours.

---

[3] « *Un système d'information est un ensemble organisé de ressources : matériel, logiciel, personnel, données, procédures, permettant d'acquérir, de traiter, de stocker des informations (sous forme de données, textes, images, sons, etc.) dans et entre des organisations.* » (Reix, 2004)
[4] Site de la plateforme 247CTF : https://247ctf.com/.
[5] Site de la plateforme Ozint : https://ozint.eu/.

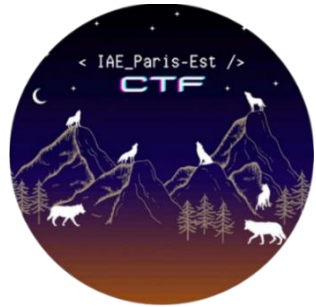

*Figure 1. Logo de l'équipe CTF pédagogique de l'IAE Paris-Est.*

| Type de compétences travaillé | Sites | Date |
|---|---|---|
| Bases dans toutes les disciplines très techniques :<br>Cryptographie<br>Réseaux<br>Intrusion<br>Retro-engineering<br>Web | https://247ctf.com/. | Juillet-Août 2023 |
| OSINT | https://ozint.eu/. | Septembre-Novembre 2023 |
| Langage Bash | https://overthewire.org/wargames | Octobre 2023 |

*Tableau 2. Planning de formation pour l'équipe CTF pédagogique (2023-2024).*

Notre équipe étant désormais prête à accompagner les membres de la formation, il a fallu mettre tout le monde au même niveau, notamment les nouveaux arrivants. Pour cela, une session découverte a été organisée pour les étudiants de premières années de licence (Figure 2), permettant ainsi d'établir un équilibre entre les niveaux.

Les éléments sur lesquels nous nous sommes concentrés se basent sur la réduction des points limitants la pratique des CTF identifiées par Szedlak et M'Manga (2020, p.3) et Vykopal *et al.* (2020, p.2). Dans un premier temps, pour remédier à (1) l'ignorance de l'existence de ce type de compétition, nous avons organisé un atelier de découverte comme indiqué précédemment. De plus, nous avons partagé une communication à ce sujet et organisé un défi CTF en 2023 pour tous les membres de la formation, avec un accompagnement important. Nous avons réalisé comme 1er CTF obligatoire pour la formation, durant toute une semaine nous avons assuré un suivi pour la réalisation du défi « Bandit » de la plateforme *Overthewire*.[7] Nous l'avons choisi, car il permet d'obtenir les bases en langage *Bash*[8] et permet de découvrir comment naviguer sur un terminal Linux, très utilisé dans les compétitions de cybersécurité. Cela a permis, de faire découvrir les défis de types *Jeopardy* à tous les étudiantes et étudiants de la formation et de leur donner les bases nécessaires à la réalisation de presque tout CTF. Tous les membres de l'équipe CTF pédagogique se sont rendus disponibles, et nous avons dû répondre en moyenne à quinze questions par jour. De plus, pour les débutants, nous avons réalisé une vidéo et une conférence pour la réalisation du tout premier niveau de ce défi afin qu'ils aient toutes les cartes en main pour commencer. Les questions posées tout au long de cette semaine ont permis la réalisation d'un *feedback* pédagogique très précis et personnalisé (Annexe 1) permettant de répondre à chaque question que se posaient les membres de la formation. Nous avons collecté des retours sur ce *feedback* qui étaient très positifs. Cependant, nous devons insister sur le contexte du CTF afin de bien délimiter les compétences travaillées du défi.

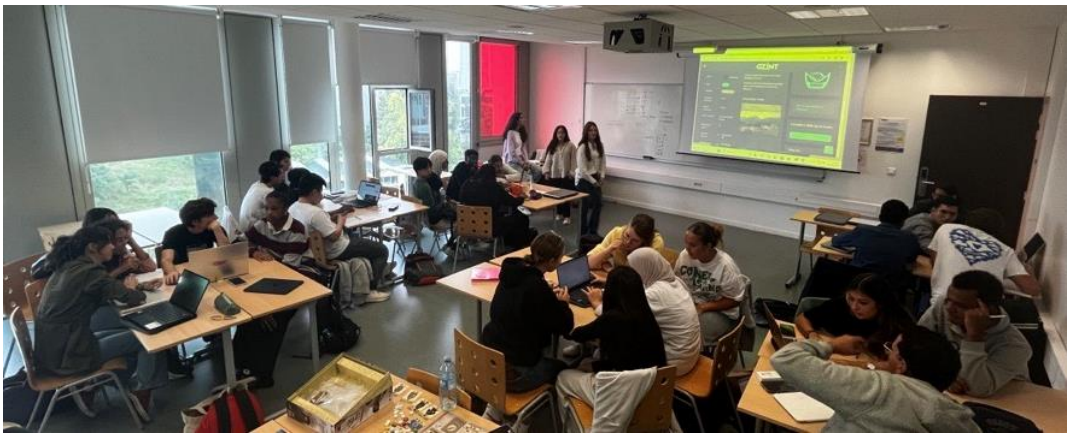

*Figure 2. Un premier CTF a été organisé dès la semaine de pré-rentrée fin août pour les nouveaux bacheliers intégrant la Licence du Parcours Informatique & Management.*

Ensuite, le point limitant la pratique des CTF sur lequel nous avons travaillé est (2) la difficulté pour les débutants à se lancer. Cela représente actuellement notre plus grand défi. Par conséquent, nous sommes très attentifs aux résultats et à l'évolution des membres de la formation. C'est la raison pour laquelle, nous avons commencé en privilégiant les CTF de type *Quizz* et *Jeopardy* afin d'attirer un large public et de faire découvrir ce qui se fait de plus courant dans le milieu sans entrer trop vite dans la technique pure. Pour cela, nous avons débuté avec un événement *Osint*, *Open Source INTelligence* (Figure 3), regroupant une cinquantaine de volontaires. Les membres de l'équipe CTF pédagogique, avaient en charge le suivi et le soutien d'une équipe de quatre personnes. Nous avons choisi ce nombre de participants par équipe, car nous avons remarqué que lorsque nous proposions des défis complexes ou longs, les participants créaient d'eux-mêmes des équipes de trois ou quatre personnes, l'inertie dans l'équipe que ce petit nombre provoque est parfait et c'est exactement ce que nous recherchons. Les participants qui terminaient les premiers recevaient un diplôme honorifique et une médaille, de plus tous les participants recevaient une récompense pour accentuer le levier de la récompense et éviter un clivage. Et cela a permis de revoir ces participants aux autres événements que nous avons organisés. À la suite de ce défi, nous avons fait remplir un questionnaire aux participants afin de récolter leur ressenti sur la difficulté, l'organisation, l'accompagnement et surtout les compétences qu'ils ont découvertes. Cela nous a permis de mettre en place un atelier de découverte d'*Osint* afin de répondre au besoin identifié à la suite de ce défi. Par ailleurs, afin d'aider les débutants, nous avons rédigé un *feedback* pédagogique assez simple et complet expliquant la logique du défi et les compétences travaillées.

---

[7] Site de la plateforme Overthewire : https://overthewire.org/wargames
[8] Langage de programmation qui permet d'exécuter des commandes sur un système d'exploitation Linux.

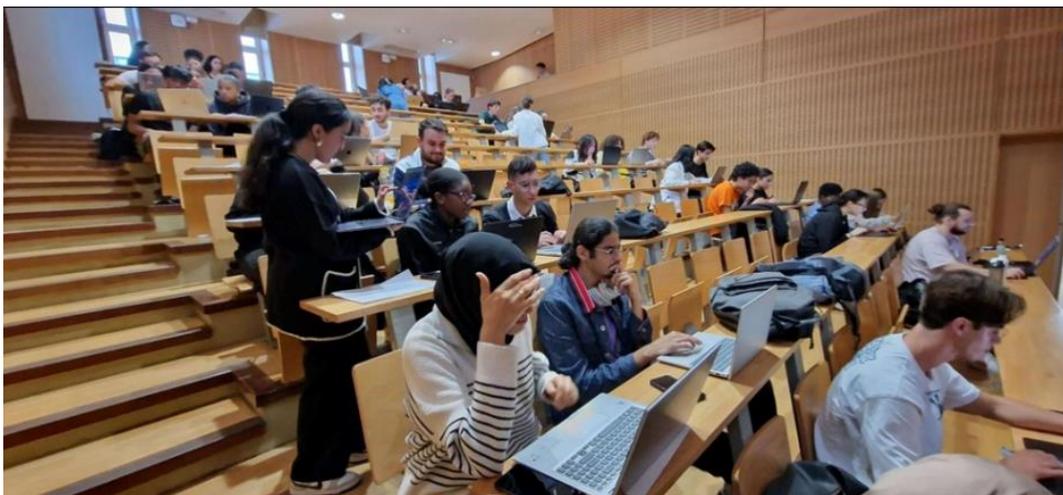

*Figure 3. Photo du premier CTF organisé par l'équipe CTF pédagogique de l'IAE Paris-Est.*

Ensuite, pour les plus curieux et volontaires de la formation qui souhaitaient pousser l'expérience plus loin, nous les avons conviés à participer à l'événement « Disparu(e)s » organisé par le collectif Oscar Zulu.[9] Nous avons communiqué sur l'événement au sein de la formation en leur apportant un soutien continu puisque les membres de l'équipe CTF pédagogique participaient également à ce défi. Nous ne donnons bien-entendu pas les *flags,* mais nous les aidons à comprendre la logique pour parvenir à les trouver. À la fin du défi, nous avons organisé une réunion avec les participants que nous avons accompagnés. Pour récolter leur ressenti et leur apporter des clarifications en plus du *write-up* que nous avons rédigé et partagé à la fin du défi. Ainsi que celui élaboré par les organisateurs. Un *write-up*, est un

---

[9] Site du défis « Disparue(s) » : https://ctf.osintisnotacrime.com/ (consulté le 5 décembre 2023)

document rédigé par les participants eux-mêmes, pour expliquer leur démarche de résolution de problèmes durant les épreuves. C'est un document qui explique le dérouler d'un CTF pour arriver au *flag*, c'est aussi un peu un journal de bord pour les participants, leur permettant de ne pas refaire les mêmes actions plusieurs et aussi de faciliter le partage d'information.

Tous ces *feedbacks* pédagogiques et *write-up* ont été mis en place dès le début pour palier au troisième et dernier point limitant la pratique des CTF, (3) l'absence de retour sur les compétitions. En réalité, ces éléments de retour après expérience se font rare, voire inexistants dans les CTF. Néanmoins, de plus en plus de *write-up* apparaissent à la fin des défis écrits par les organisateurs comme nous le démontre la plateforme Ozint.eu.[10] ou encore les défis proposés par l'ANSSI avec l'Hackropole.[11] par exemple. Du côté de l'équipe CTF, afin de pallier ce problème, nous avons, sous la supervision d'un enseignant, rédigé des comptes-rendus pédagogiques détaillés des défis, avec des explications précises des techniques utilisées et de la logique attendue pour les résoudre. Ce sont des *writes-up* complets qui s'appuie sur la littérature scientifique afin d'expliquer le plus clairement et objectivement possible le contexte et les compétences travaillés. Cette utilisation de la revue scientifique dans nos comptes-rendus pédagogiques détaillés permet d'identifier aussi les prochaines étapes à suivre de la part de notre équipe afin d'aider au mieux les étudiantes et étudiants à développer leurs connaissances.

En plus de la résolution de ces trois point limitant la pratique des CTF, notre équipe CTF pédagogique participe à des événements réputés afin de progresser et de voir ce qu'il se fait ailleurs, aux côtés des meilleurs comme le Purple Pills Challenge.[12] Nous avons, pour maintenir cette progression au sein de l'équipe et des événements que nous proposons, mis en place des ateliers d'échanges et de *master class*. En d'autres termes, nous échangeons sur des défis complexes, et parfois un membre de l'équipe nous explique comment il a résolu un défi difficile. C'est ce que nous avons fait, par exemple avec certains défis de la plateforme 247CTF et plus particulièrement pour les défis de BattleH4ck.[13] Nous avons étendu cette idée d'atelier pour la progression des membres de la formation. Dans ce cas-ci, l'objectif est de faire découvrir des outils, des techniques ou parfois des méthodes de travail. Aussi, des cours d'explication de résolution de défis complexe, rendus accessibles, afin de pousser l'accompagnement à son maximum. C'est ce dernier type d'événement qui nous est le plus demandé puisque l'accompagnement est maximal. Nous avons même planifié des cours pour la résolution de défis CTF particuliers. Vous pourrez revoir la planification de tous les événements que nous avons organisé avant février 2024 dans l'annexe 2.

---

[10] Site de la plateforme Ozint.eu : https://ozint.eu/.
[11] Site de la plateforme Hackropole : https://hackropole.fr/fr/challenge/.
[12] Site du *Purple Pill Challenge* : https://www.purplepilldéfis.fr/.
[13] Site du Battlehack : https://seela.io/battleh4ck/.

Lors de ce projet nous avons remarqué que le niveau des participants, ne dépendait pas de leur âge ou de leur niveau d'étude mais surtout des *softs skills* et parmi nos premiers résultats nous avons pu le remarquer. « *Soft skills correspond the human personal characteristics and responsible for their social adaptation in the work group.* » (Gurjanov *et al.*, 2020)

## 4. La méthodologie et les premiers résultats visibles

Nous avons déjà effectué des expérimentations lors du premier semestre et les premiers résultats nous servent de fondations pour continuer ce travail de recherche-action (Chaliès, 2018) durant le second semestre de cette année universitaire 2023-2024. Pour obtenir ces résultats, nous avons fonctionné de la manière suivante :

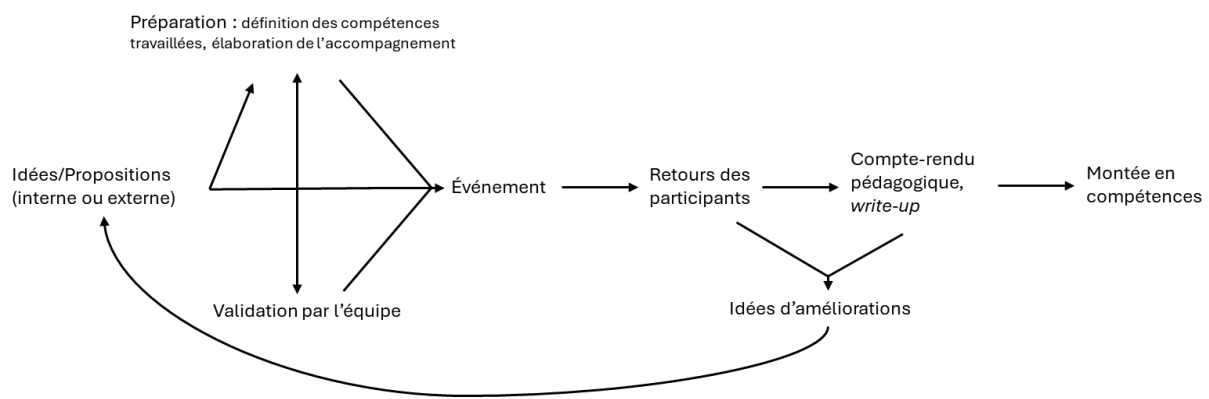

*Figure 4. Schémas de l'organisation de la méthodologie de recherche-action de l'équipe CTF pédagogique (2023-2024).*

Nous commençons par échanger sur le défi que nous souhaitons proposer à la formation et son format, un événement compétition, présentation, cours. Et de toutes les caractéristiques de l'événement : durée, date, obligatoire ou non, public visée, etc… Après cela, nous préparons le suivi, tout simplement en faisant l'activité chacun de notre côté et en réalisant tous un *write-up*, que nous mettons en commun pour rédiger le *write-up* finale, que nous partagerons à la formation à la fin de l'événement. Ce document doit être validé par l'équipe complète. Ensuite, nous organisons l'événement et d'après les retours des participants sur l'événement et sur le *feedback* pédagogique et le *write-up,* nous ciblons les éléments d'amélioration et renouvelons le processus. À cela, s'ajoute l'étude de la montée en compétences des participants, cette étude est sur le long terme, au fur et à mesure des défis nous élaborons des éléments de mesure pour les compétences. Cela représente un élément de résultat que nous n'avons pas encore pu définir complétement.

Le premier résultat que nous avons constaté est le nombre optimal de membres lors de la formation des équipes, que nous avons évoqué précédemment. Lorsque nous laissions le choix à nos participants de former des équipes, ils se mettaient toujours par équipe de trois ou quatre. Ensuite, nous avons remarqué que les équipes de plus de six personnes demandaient une communication plus complexe et des compétences de management plus importantes. Dans ces équipes, quatre personnes maximums peuvent réellement coopérer et participer aux défis ensemble.

Le niveau et l'âge n'ont rien avoir avec le niveau de compétence, c'est l'intérêt porté à la discipline et l'investissement qui font la différence, nous y reviendrons plus tard. C'est pour cela qu'il faut créer plusieurs types d'événement, adapté à chaque niveau :

- Pour les débutants qui seraient réticents de se lancer, on utilise plus des défis type *Quizz* ou *Jeopardy* avec des explications à chaque question.
- Pour les débutants/intermédiaires qui se sont lancés, mais n'ont pas les fondamentaux techniques, on utilise des défis *Jeopardy* avec des questions spécialisées sur les techniques de CTF.
- Pour les intermédiaires qui ont les fondamentaux techniques, on utilise des défis *Jeopardy* poussé, utilisant les compétences techniques se rapprochant de plus en plus du type *Attack-Defense*.
- Pour les intermédiaires/avancés, qui ont les fondamentaux techniques et un véritable intérêt pour la discipline, on utilise les défis de types *Attack-Defense* et *King of the Hill*.
- Pour les avancés, qui souhaitent accomplir des performances et participer à des compétitions, on les accompagne dans la participation de CTF régionale voire nationale.

Déterminer ces différents niveaux a pris du temps et nous a demandé d'identifier trois facteurs : [1] les connaissances techniques en *hacking*. [2] L'investissement et le niveau d'intérêt apportés à ce domaine. [3] La capacité de compréhension et d'adaptation aux défis. Grâce à cela et à nos premières expérimentations, qui étaient suivi par des questionnaires, nous avons pu établir un spectre de l'état actuel du notre formation (Figure 5). Les résultats de ce spectre proviennent à la fois des questionnaires et des résultats obtenus lors des CTF que nous avons organisés. Ceci va nous servir de base pour atteindre le maximum de personnes au niveau intermédiaire à la fin de l'année, nous pouvons voir qu'il reste environ la moitié des membres de la formation à accompagner pour atteindre ce niveau. En effet, nous avions remarqué aux débuts que certes les compétences techniques étaient importantes pour commencer dans la discipline, mais pour avancer, il fallait plus. Et nous remarquons que, malgré les différences de niveau d'étude, la répartition est assez homogène.

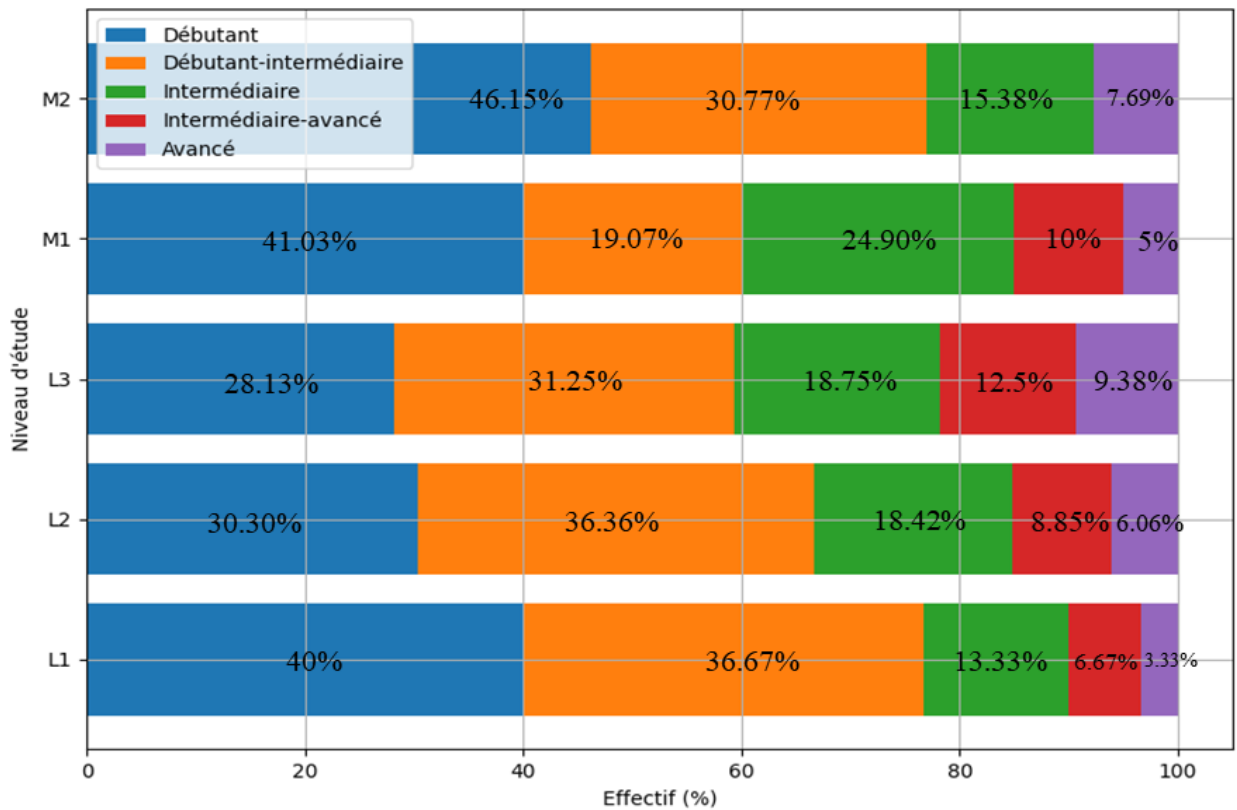

*Figure 5. Spectre de compétence en fonction du niveau d'étude au sein de la formation Informatique et Management de l'université Paris-Est.*

Les étudiantes et étudiants de master n'ont pas d'avance sur les licences, donc ces *soft skills* jouent un rôle crucial dans cette montée en compétences, puisque les connaissances techniques acquises en cours ne suffisent pas. Nous n'écartons pas l'hypothèse que certaines personnes ont des facilités pour appréhender et comprendre cet aspect technique de la cybersécurité. Mais selon nos enquêtes et les membres de l'équipe CTF pédagogique, ces facilités ne sont pas suffisantes pour gagner une compétition. Ces *soft skills* sont nécessaires pour accomplir une performance et progresser. Lorsqu'on les aborde, nous mettons principalement l'accent sur les compétences suivantes : la détermination, l'investissement personnel, la communication, l'esprit d'équipe et l'intégrité. La détermination est particulièrement cruciale pour les débutants, car le domaine initial est si vaste qu'il faut une forte volonté pour persévérer.

La communication et l'esprit d'équipe sont également primordiaux, étant donné que le travail se fait souvent en collaboration, impliquant le partage des connaissances et des idées. Comme en témoignent les résultats de deux défis. Dans le premier défis *Osint* proposé lors de la formation, les deux premières places ont été remportées par des équipes très collaboratives (Segal & Gerstel, 2019), tandis que trois autres équipes individualistes (Kainan, 1992), se classaient troisième, cinquième et septième sur douze. Le même schéma s'est reproduit avec le défi « Bandit » présenté précédemment. De plus, lors de la conception des défis, l'équipe CTF pédagogique a toujours la charge de les tester. Bien que nous travaillions individuellement pour tester autant de défis que possible, nous rencontrons souvent des difficultés trop importantes pour les résoudre seuls. Nous devons alors solliciter l'aide des autres membres de l'équipe CTF pédagogique pour progresser. Ainsi, il est clair qu'avec trois, quatre personnes, nous sommes plus efficaces et nous allons plus loin que si nous étions seuls. Nous n'avons pas encore suffisamment de résultats pour confirmer cette dernière affirmation, ce qui constitue également l'un de nos objectifs dans la poursuite de ce projet.

## 5. Conclusion

L'intégration des compétitions informatiques, notamment les *Capture The Flag* (CTF), dans la formation en informatique et management de l'IAE Paris-Est et son master de Management de la Sécurité des Système d'Information représente une approche novatrice et efficace pour développer les compétences des étudiantes et étudiants en cybersécurité. Notre projet de formation, axé sur l'équipe CTF pédagogique, vise à offrir une expérience à la fois ludique et pratique. En proposant des séances de découverte et des défis adaptés, nous favorisons la progression d'apprentissage pour tous les participants, quel que soit leur niveau initial. Les événements CTF suscitent l'intérêt des étudiantes et étudiants et leur offrent un soutien personnalisé pour surmonter les difficultés. Nos premiers résultats montrent une corrélation entre l'investissement personnel, l'acquisition de compétences techniques et le développement des *soft skills*. En explorant de nouvelles voies telles que les *escape games* axés sur les systèmes d'information, nous envisageons d'élargir notre méthodologie afin de rendre les concepts de sécurité informatique plus tangibles et accessibles à un public plus large. En effet, nous avons commencé à expérimenter une nouvelle approche plus axée sur le système d'information que sur le système informatique[14], en utilisant des *escape game*. Cette idée a pour objectif de nous

---

[14] « *Un système informatique est un ensemble de moyens informatiques et de télécommunications, matériels et logiciels, ayant pour finalité de collecter, traiter, stocker, acheminer et présenter des données*. » (Lonchamp, 2017)

éloigner un peu de la technique pour mieux exploiter ces *soft skills* et ainsi pouvoir atteindre un public encore plus large et les aider à mettre un pied dans ce domaine. Ainsi on a un moyen de continuer à combler le second point limitant la pratique des CTF. Nous avons donc conçu deux activités réalistes dans lesquelles les participants font de l'ingénierie sociale.[15] et en tire profit. Dans ce type d'activité, nos premiers résultats confirment l'importance des *soft skills* qui facilite le travail d'équipe. Seul le champ de compétences, en fonction du niveau d'études, est amené à évoluer, car il est moins crucial d'avoir des connaissances techniques dans ce domaine, le rendant ainsi plus accessible. De plus, s'intéresser aux *mad skills*, qui est un concept qui correspond aux compétences rares, assez folles, aux singularités qui sont au service de l'innovation (Lamri *et al.*, 2022) semble représenter un axe d'étude pertinent sur ce sujet. Dans l'ensemble, notre projet démontre l'efficacité d'une approche pratique et ludique pour former la prochaine génération d'étudiantes et étudiants en cybersécurité, en comblant les lacunes éducatives et en favorisant une meilleure compréhension des enjeux liés à la sécurité des systèmes d'information.

## 6. Références

---

[15] « *L'ingénierie sociale est une manipulation consistant à obtenir un bien ou une information, en exploitant la confiance, l'ignorance ou la crédulité de tierces personnes* » (ANSSI, 2019 ; Penven, 2013)

# 7. Annexes

Annexe 1. Exemple de compte-rendu pédagogique

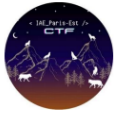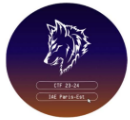

Annexe 2. Tableau des activités organisées par l'équipe CTF pédagogique

| Dates | Activités proposées |
| --- | --- |
| 31 août 2023 | Challenge découverte |
| 28 octobre 2023 | Premier défi CTF (Osint) |
| 8 octobre – 16 octobre 2023 | Defi Bandit |
| 18 octobre 2023 | Participation au Purple Pills Challenge |
| 01 novembre – 30 novembre 2023 | Participation à l'événement « Osint is not a crime » du collectif Oscar Zulu |
| 22 novembre 2023 | Défi CTF pour les Master 2 (247CTF) |
| 16 décembre 2023 | Participation à l'événement Oteria Cyber Cup |
| 19 décembre 2023 | Cours, découvertes des outils OSINT |
| 25 janvier 2024 | Escape Game |
| 15 février 2024 | *Master class* interne d'un défi BattleH4ck |
| 29 février 2024 | CTF machines virtuelles |
| 19 mars | Conférence organisée pour la CASDEN sur l'Osint |
| 27 mars | Cours sur les machines virtuelles et les bases pour un CTF |